\documentclass{moriond}
\usepackage{amssymb}

\def\Journal#1#2#3#4{{#1} {\bf #2}, #3 (#4)}

\def\PRD{{\em Phys. Rev.} D}
\def\jcap{\em J.~Cosmol.~Astropart.~Phys.}
\def\mnras{\em Mon.~Not.~R.~Astron.~Soc.}
\def\apjs{\em Astrophys.~J.~Supp.}
\def\physrep{\em Phys.~Rep.}

\def\be{\begin{equation}}
\def\ee{\end{equation}}
\def\bea{\begin{eqnarray}}
\def\eea{\end{eqnarray}}

\newcommand{\Photo}{\includegraphics[height=35mm]{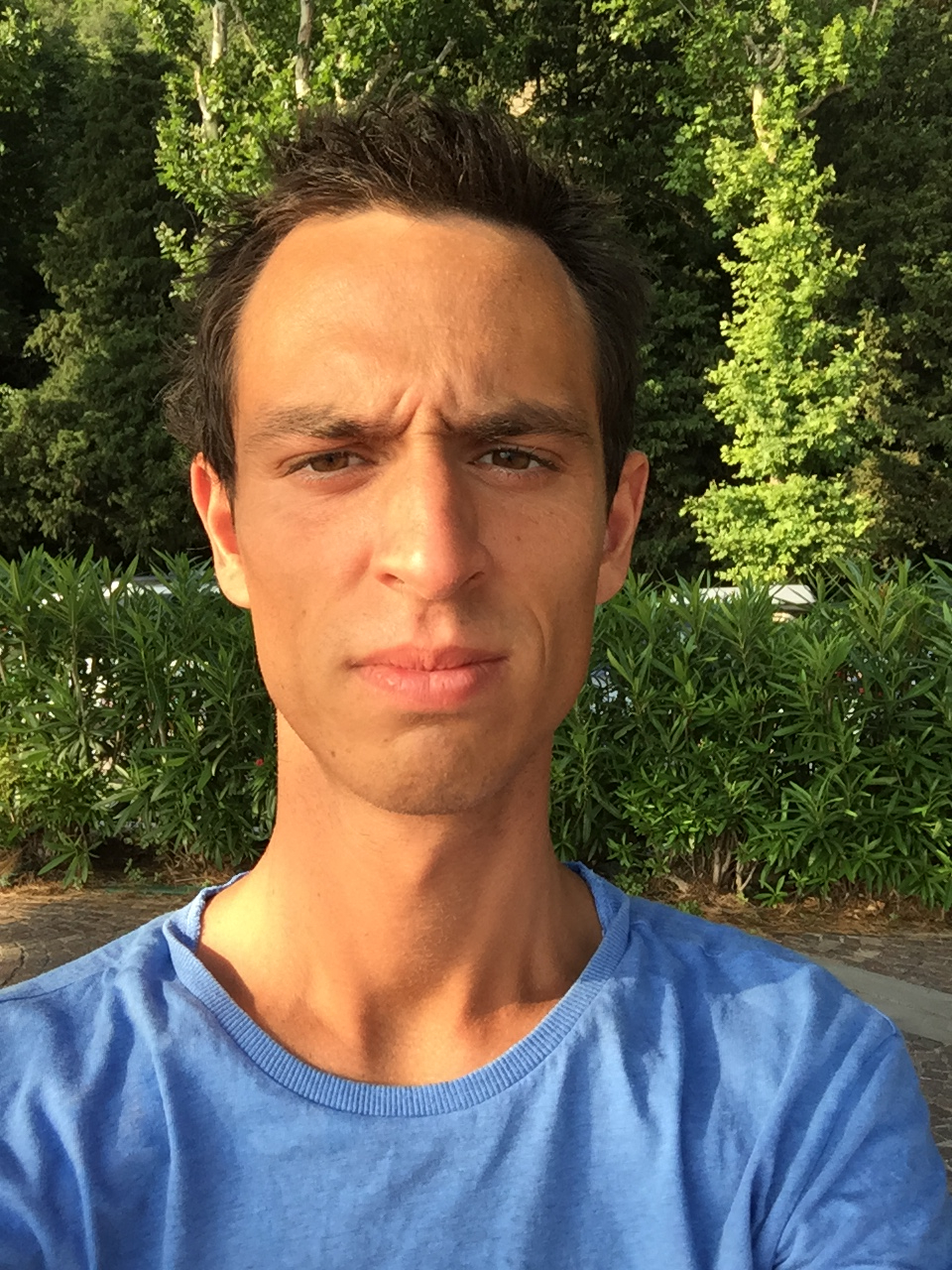}}

\begin{document}
\vspace*{4cm}
\title{MODELLING THE MATTER BISPECTRUM TOWARDS NONLINEAR SCALES - TWO AND THREE LOOPS IN PERTURBATION THEORIES}

\author{ A. LAZANU }

\address{INFN, Sezione di Padova,
via Marzolo 8, I-35131, Padova, Italy}

\maketitle\abstracts{
I compute the matter bispectrum of large-scale structure up to two loops in the Standard Perturbation Theory and up to three loops in the \textsc{MPTbreeze} renormalised perturbation theory, determining the contributing loop diagrams and evaluating them numerically. In the process I remove the leading divergences in the integrands, thus making them infrared-safe. By comparing the results to numerical simulations, I show that in the case of the Standard Perturbation Theory, the bispectrum at two loops is more accurate than at one loop, up to $k_{\textrm{max}} \sim 0.09 \, h/\textrm{Mpc}$ at $z=0$ and $k_{\textrm{max}} \sim 0.11 \, h/\textrm{Mpc}$ at $z=1$.  The \textsc{MPTbreeze} can be employed to accurately model the matter bispectrum up to $k_{\textrm{max}} \sim 0.17 \, h/\textrm{Mpc}$ at $z=0$ and $k_{\textrm{max}} \sim 0.24 \, h/\textrm{Mpc}$ at $z=1$ using the results at three loops.
}
\section{Introduction}
The late-time galaxy distribution in the Universe is providing an increasing amount of information from large-scale structure (LSS) surveys. In particular, correlations in Fourier space of the galaxy overdensity can be used to test the standard cosmological model and to explore alternatives and to look for possible new features. In order to achieve this, one requires an accurate modelling of the corresponding quantities for dark matter, together with the nonlinear relationship between the galaxy and dark matter distributions (bias). Large scales are linear and well modelled, but smaller scales are increasingly nonlinear and less understood and it is where new cosmological features may be hidden. 

In this work, I am investigating the matter three-point correlation function (bispectrum) of LSS for Gaussian initial conditions and in particular I analyse methods to study nonlinearities in an analytic fashion using perturbation theories at more than one loop. These methods are desirable, since the alternative, using $N$-body simulations, is computationally expensive. I determine the matter bispectrum in the Eulerian Standard Perturbation Theory (SPT) up to two loops and in the renormalised perturbation theory (\textsc{MPTbreeze}) up to three loops \cite{Lazanu2018}. The calculation of these bispectra presents significant challenges from both a theoretical and numerical point of view: there are many terms to be considered, with large amplitudes and alternating signs, which contain divergences that must be removed and which involve integrals in many dimensions (six at two loops and nine at three loops). 
	
\section{The matter bispectrum in perturbation theory}
\subsection{Eulerian Standard Perturbation Theory}
The SPT represents the most straightforward extension to linear theory and its basic features can be derived by considering small departures from the homogeneous expansion of the Universe, and deriving coupled evolution equations for the matter overdensity ($\delta$) and for the divergence of the velocity ($\theta$) in Fourier space \cite{Bernardeau2002}. By considering expansions of these two quantities, solutions can be derived at any order, which can be employed to determine correlation functions. This perturbative scheme is however not convergent, with higher loop orders not guaranteeing a better accuracy. 
In this work I determine analytically the bispectrum at two loops in SPT. There are 11 terms (each containing a number of permutations) (see Ref. \cite{Lazanu2018} for the full list). 

\subsection{Renormalised Perturbation Theory}
The renormalised perturbation theory \cite{Crocce2006} has been developed to cure some of the problems in SPT. The dark matter fluid equations are solved using nonlinear propagators, in order to derive a resummed perturbative expansion that is also convergent. Its simplification, \textsc{MPTbreeze} \cite{Crocce2012}, provides a framework to derive correlation functions using generating functions \cite{Bernardeau2008}. In particular, the bispectrum can be expressed in terms of a selection of SPT terms, overall multiplied by a decaying exponential function which depends on the linear matter power spectrum. In this work I present predictions of this theory at two and three loops for the bispectrum \cite{Lazanu2018}, extending and improving previous results \cite{Lazanu2016,Lazanu2017,Lazanu2017b}.

\subsection{IR-safe integrands and numerical implementation}
At two and three loops, the integrals that must be calculated are six and nine-dimensional respectively. The corresponding integrands contain infrared divergences and to increase the accuracy of the results, I exploit the symmetries and infinite domain of integration to remove the leading divergences by moving them to a finite number of points \cite{Carrasco2014,Angulo2015}. After applying this procedure, I employ the  Monte-Carlo integrator \textsc{Cuba} \cite{Hahn2005} for the numerical part of the work.

\section{Comparison with simulations}
\subsection{Simulations}
For the comparison, I am employing Gaussian simulations \cite{Schmittfull2013}, with three realisations, containing $512^3$ particles in a box size of $1600\, \mathrm{Mpc}/h$, yielding a wavevector range of $[0.0039, 0.5]\, h/\mathrm{Mpc}$. They assume 2LPT initial conditions and are evolved from $z=49$ until today using the \textsc{Gadget}-3 code \cite{Springel2005}. The cosmology considered is a flat $\Lambda$CDM universe with WMAP7 parameters \cite{Komatsu2011}.

\begin{figure}
\centering
\includegraphics[width=0.87\linewidth,trim={1.14cm 0.48cm 2.02cm 1.05cm},clip]{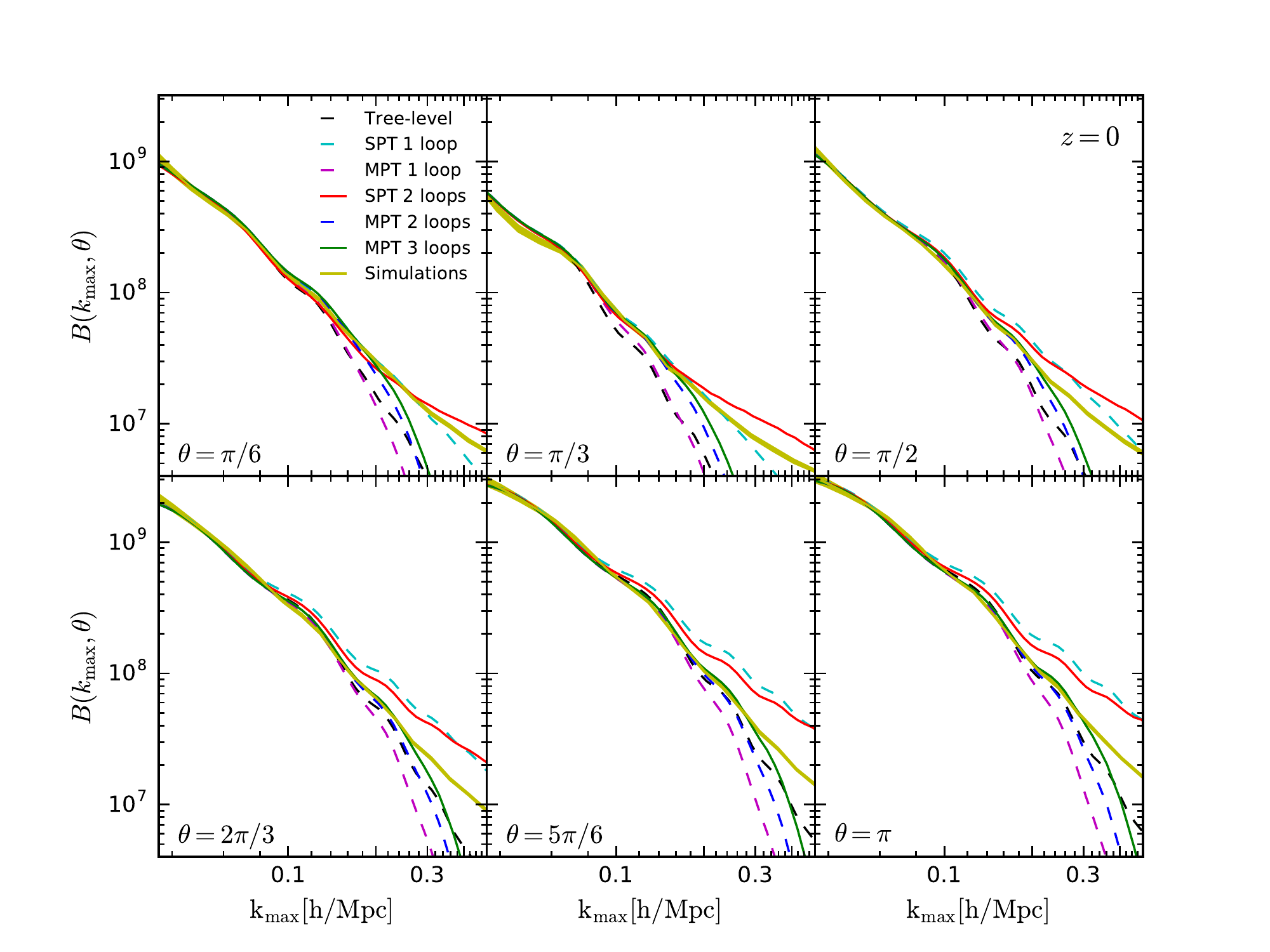} 
\includegraphics[width=0.87\linewidth,trim={1.14cm 0.48cm 2.02cm 1.05cm},clip]{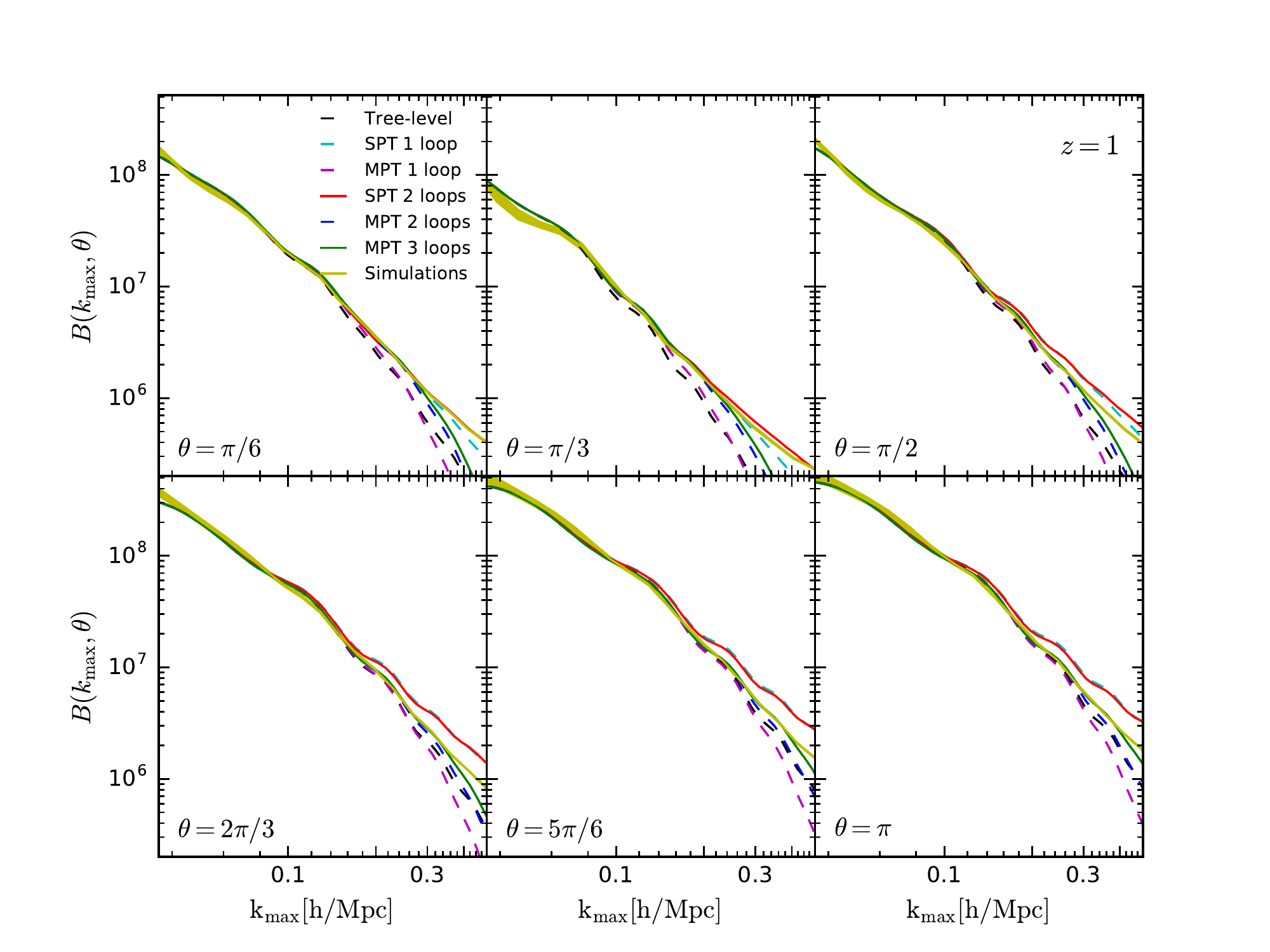}
\label{fig:z01}
\caption[]{The bispectra in perturbation theories at $z=0$ (top) and $z=1$ (bottom) for isosceles triangles, represented in terms of the maximum side of the triangle and the angle between the equal sides. The perturbative methods considered are the SPT up to two loops and the \textsc{MPTbreeze} up to three loops. Simulations are plotted in yellow for comparison.}
\end{figure}

\subsection{Results}
In this section, I show the performance of the two perturbation schemes discussed by evaluating and plotting the matter bispectrum in six triangle configurations (Figure 1) at redshifts $z=0$ (top) and $z=1$ (bottom) in order to check on which scales each of them is accurate. The bispectra are plotted for isosceles triangles, in terms of the largest triangle side and the angle between the equal sides. Going from left to right and top to bottom, the triangles considered vary from squeezed to flattened. The two-loop SPT and three-loop \textsc{MPTbreeze} bispectra are plotted as continuous lines, and the others as dashed lines. The simulations are represented as a yellow band, with a thickness representing the error bars from the three realisations. 

In all the configurations considered, the SPT at one loop provides excessive signal on mildly nonlinear scales \cite{Bernardeau2002}. At two loops, this excess is diminished on the scales of interest, providing a better fit to the simulations in the quasi-nonlinear regime, before  diverging on smaller scales. Even so, it only provides  adequate predictions for scales $\max(k_1,k_2,k_2) \lesssim 0.09 \, h/\mathrm{Mpc}$ at $z=0$ with a modest improvement to $\max(k_1,k_2,k_2) \lesssim 0.11 \, h/\mathrm{Mpc}$ at $z=1$. The two-loop SPT bispectrum matches simulations to smaller scales in squeezed configurations, with a worsening of the fit as the angle is increased.

For \textsc{MPTbreeze}, as a convergent expansion, it can be confirmed from the plots that going from one to two loops provides a significant improvement in the range of validity of the theory, while adding further the three-loop terms yields a smaller gain. Thus, this method adequately describes simulations for scales $\max(k_1,k_2,k_3) \lesssim 0.08 \, h/\mathrm{Mpc}$,  $\max(k_1,k_2,k_3) \lesssim 0.14 \, h/\mathrm{Mpc}$ and $\max(k_1,k_2,k_2) \lesssim 0.17 \, h/\mathrm{Mpc}$ at one, two and three loops respectively at $z=0$. At $z=1$, this is increased to $0.11 \, h/\mathrm{Mpc}$, $0.19 \, h/\mathrm{Mpc}$ and $0.24 \, h/\mathrm{Mpc}$ respectively. In both cases, the tightest constraints appear in the equilateral configuration, with matches up to significantly smaller scales in flattened configurations.

As a further check of these conclusions, I have employed a different set of simulations \cite{Sefusatti2010} using a slightly different cosmology, and I have shown that the results are indeed robust \cite{Lazanu2018}.

\section{Conclusions}
In this work, I have shown that a better agreement with simulations can be obtained for the matter bispectrum by going to a higher loop order in perturbation theories. In particular, \textsc{MPTbreeze} can be employed to accurately describe the matter bispectrum of LSS on mildly nonlinear scales, up to scales $k_{\mathrm{max}} \sim 0.14 \, h/\mathrm{Mpc}$ and $k_{\mathrm{max}} \sim 0.17 \, h/\mathrm{Mpc}$ at two and three loops respectively at $z=0$. The improvement is obtained however at a significant additional computational cost. For the SPT at two loops, there is only a small improvement with respect to the one loop result, but this facilitates further developments in the field that may significantly increase the scales of validity of the method (e.g. the computation of the two-loop bispectrum in the EFTofLSS).

\section*{Acknowledgements}

I am grateful to Marcel Schmittfull for sharing the measurements of the bispectra for the simulations used in the comparison and to Paul Shellard for useful discussions in the early stages of the development of this work. I would also like to thank the organisers for creating a stimulating environment in the {\it 53rd Rencontres de Moriond}.

\end{document}